# Unfolding femtoscale ionic movement in CuO through polarized Raman spectroscopy

Binoy Krishna De, Vivek Dwij and V.G. Sathe*
UGC-DAE Consortium for Scientific Research,
University Campus, Khandwa Road, Indore-452001, INDIA
*vasant@csr.res.in

**Abstract:** Recently, CuO has been proposed as a potential multiferroic material with high transition temperature. Competing models based on spin current and ionic displacements are invoked to explain ferroelectricity in CuO. The theoretical model predicting ionic displacement suggested that the shift in ions is essentially along *b*-axis with very small amplitude (~$10^{-5}$ Å). Experimentally detecting displacements of such a small amplitude in a particular direction is extremely challenging. Through our detailed polarized Raman spectroscopy study on epitaxial film of CuO, we have validated the theoretical study and provided direct evidence of displacement along the *b*-axis. Our study provides important contribution in the high temperature multiferroic compounds and showed for the first time, the use of the polarized Raman scattering in detecting ionic displacements at the femto-scale.

## Introduction

Magnetically induced ferroeletricity in a single phase material is a topic of frontline reaserch due to its vast potential applications. Because of strong cross coupling between the electric and the magnetic properties in this type of materials, electric polarization can be controlled by tuning the magnetic field or vice versa. The main hurdle in the applications of these materials is the low ferroelectric transition temperature, $T_C$ (typical $T_C$ < 40K) shown by these materials. The ferroelectricity in such materials arises due to the spiral magnetic order which essentially possesses competing magnetic interactions and thus lowers the magnetic ordering temperature. In this respect, CuO is attracting attention due to its unconventional high Tc. It shows two magnetic transitions, one at 230K ($T_{N2}$) (*1*) where the spins show non-collinear incommensurate order with modulation vector along (0.006, 0, 0.017) direction (*2*) and the other at 213K ($T_{N1}$), below which collinear commensurate antiferromagnetic spin order appears (*1*). In collinear commensurate ordered state, spins are aligned along the crystallographic '*b*' direction (*3*). In the temperature window between $T_{N1}$ and $T_{N2}$, electrical polarization appears along the *b*-axis through Dzialoshinsky-Moria (D-M) interaction (*4, 5*). Jin *et al* (*4*) and Giovannetti *et al* (*5*) attempted to explain the possible origins of such high temperature ferroelectricity using DFT calculations. Jin *et al* made a definitive prediction of ionic displacements responsible for the induced ferroelectricity arising from the D-M interaction in the non-collinear spin spiral state (*4*) while Giovannetti *et al* attributed the ferroelectricity to the variation in electronic density distribution because of the spin current induced due to spiral magnetic order (*5*). Jin *et al* calculated the amplitude of possible ionic displacements to be ~$7 \times 10^{-5}$ Å along *b*-axis corresponding to the experimentally reported dipole moment (*2*). Therefore, in order to understand the root cause of ferroelectricity shown by CuO, it is important to experimentally probe the ionic displacement. However, detection of such small ionic displacement (in femto-scale order) is extremely difficult. Walker et al (*6*) detected indirectly the femtoscale atomic displacements in TbMnO$_3$ using resonant magnetic X-ray scattering. Here, we propose Angle Resolved Polarized Raman Spectroscopy (ARPRS) to be an efficient tool to detect small ionic displacements in a particular direction. According to the basic principles of Raman spectroscopy, even very small ionic displacement should change the specific Raman tensor element through the change of polarizability derivative associated with the displacement along a particular direction. The Raman intensity is directly proportional to the square of the Raman tensor elements. Thus, measuring the Raman intensity associated with the proper tensor elements should empower one to trace very small atomic displacements.

In this report, the temperature dependent ARPRS study of CuO epitaxial thin film is presented through which we are able to estimate the direction and magnitude of the atomic displacements across the commensurate to incommensurate magnetic order. Raman tensor elements are calculated by fitting the intensity variations with respect to the various in-plane rotation angles using theoretically calculated equations. Obtained elements are used to probe the lattice displacements in magnetically ordered state. Detail ARPRS studies provide a direct experimental evidence of ionic displacements which is the dominate cause of ferroelectricity in CuO as proposed by Jin *et al* through theoretical calculations (*4*).

**Results and Discussions:**

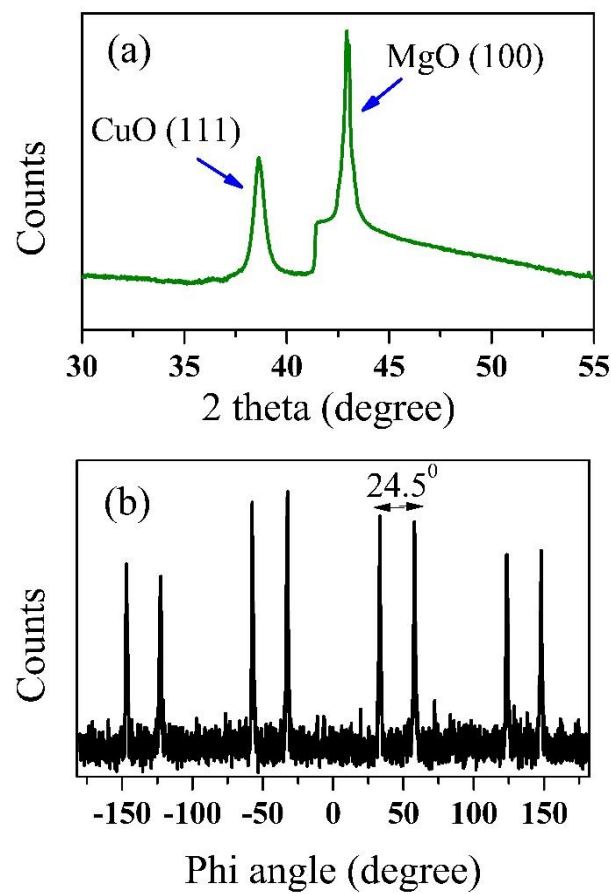

**Fig 1:** (a) X-ray θ-2θ scan on the CuO film grown on MgO (100) single crystal substrate. (b) Azimuthal φ-scan on the ($\bar{1}$11) reflection of the CuO film.

The film was deposited using pulsed laser deposition and thoroughly characterized using x-ray diffraction studies and found to be epitaxial in nature along the (111) direction (fig 1(a) and (b)). The details of the thin film preparation and characterization is given in supplementary material (*7*). The film was found to be tensile strained in the out-of-plane direction which is expected to change the magnetic ordering temperature. The magnetization measurements as a function of temperature showed enhancement in the magnetic ordering temperature ($T_{N2}$) from 230 K to 252 K (supplementary figure S1). The magnetization studies failed to detect the incommensurate to commensurate transition ($T_{N1}$). It was previously reported that the application of pressure enhances the $T_{N2}$ of CuO (*8*). The presence of out-of-plane tensile strain in our sample suggests that the film has in-plane compressive strain. Further confirmation of the in-plane compressive strain and epitaxial relationship was obtained from the off-specular azimuthal φ-scans which were carried out on CuO ($\bar{1}$11) and (202) planes. The stereographic projection of the growth direction (111) and ($\bar{1}\bar{1}\bar{1}$) do not superimpose on each other leading to two families of peaks in the φ-scan (*9*). This results in 8 peaks in the φ-scan, the typical φ-scan on CuO ($\bar{1}$11) is shown in figure 1(b). Numerical calculation suggests that the angle between the two successive peaks in the φ-scan should be 21.68° (see supplementary materials for details (*7*)), whereas the experimentally observed values are 24.5°. We attempted to back calculate the lattice parameters using this angular separation and it resulted in significant compression of *a*, *c* and *β* while a minimal elongation in *b* lattice parameter. The room temperature Raman modes when compared with the bulk compound also provided similar results [details of the strain calculations is given in supplementary material (*7*)].

As mentioned before, the direction and magnitude of the atomic displacement in this compound in multiferoelectric phase is predicted to be along the *b*-axis. In order to detect this, detailed polarization dependent Raman spectroscopy study was under taken on the film. In the present experimental geometry, the film is oriented in the (111) direction which is thus parallel to the incident and scattered beam. The film was mounted on a rotation stage and Raman spectra were collected as a function of in-plane rotation angle ($\psi$). The polarization plane of the incident laser beam is thus allowed to rotate with an angle with the axis of the unit cell and the polarization of the scattered beam was analysed using an analyser. In order to estimate the direction and magnitude of the displacement of Cu and O ions, the Raman tensor elements were first estimated.

The group theoretical calculations on monoclinic CuO (*10*) ($C_{2h}$ with 2 molecule per primitive unit cell) results in twelve vibrational modes, given as: $\Gamma_{total}=4A_u+5B_u+A_g+2B_g$ (*11*). Raman selection rule implies that only $A_g+2B_g$ modes are Raman active (*11*). $A_g$ mode located at 298cm$^{-1}$ (*12*) corresponds to vibration of oxygen along the crystallographic *b*-axis (*13*) while $B_{1g}$ mode (345 cm$^{-1}$) corresponds to oxygen vibration along *a*-axis and $B_{2g}$ (632cm$^{-1}$) (*13*) represents vibration of the oxygen perpendicular to both *a*- and *b*-axes. Raman tensors of $A_g$ and $B_g$ modes are represented as: (*11*)

$$A_g = \begin{pmatrix} a & 0 & d \\ 0 & b & 0 \\ d & 0 & c \end{pmatrix} \quad (1)$$

$$B_g = \begin{pmatrix} 0 & e & 0 \\ e & 0 & f \\ 0 & f & 0 \end{pmatrix} \quad (2)$$

$a, b, c, d, e$ and $f$ are Raman tensor elements. For the CuO epitaxial film grown in (111) direction, the intensity variation of the $A_g$ and $B_g$ modes with in-plane rotation angle $\psi$ in parallel and cross polarization configuration can be given as (See supplementary material for details (*7*)):

$$I_{A_g}^{\parallel} \propto \left((0.64a+0.34b)cos^2\psi + (0.15a+0.28b+0.54c-0.56d)sin^2\psi + (0.31(a-b)-0.58d)sin2\psi\right)^2 \quad (3)$$

$$I_{B_g}^{\parallel} \propto \left((-0.95e)cos^2\psi + 2(0.21e-0.20f)sin^2\psi + (0.20e+0.43f)sin2\psi\right)^2 \quad (4)$$

$$I_{A_g}^{\perp} \propto \left((0.31(b-a)+0.58d)cos2\psi + \frac{(0.49a+0.04b-0.54c+0.56d)}{2}sin2\psi\right)^2 \quad (5)$$

$$I_{B_g}^{\perp} \propto \left((-0.21e - 0.43f)\cos 2\psi + (-0.69e + 0.42f)\sin 2\psi\right)^2 \quad (6)$$

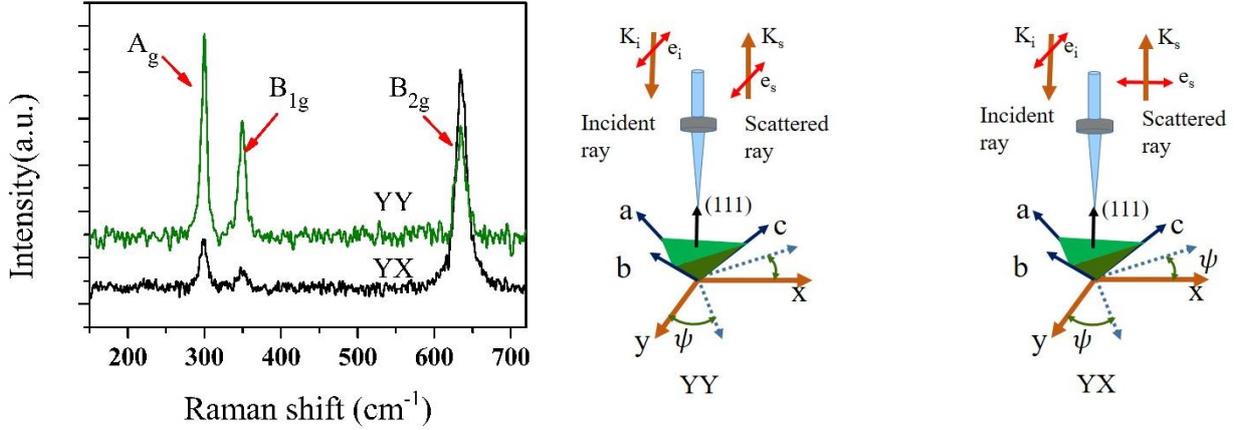

**Fig 2**: Raman spectra collected in parallel (YY) and cross (YX) polarization configuration on CuO epitaxial film with fixed $\psi$. The panel at the right hand shows the experimental geometry.

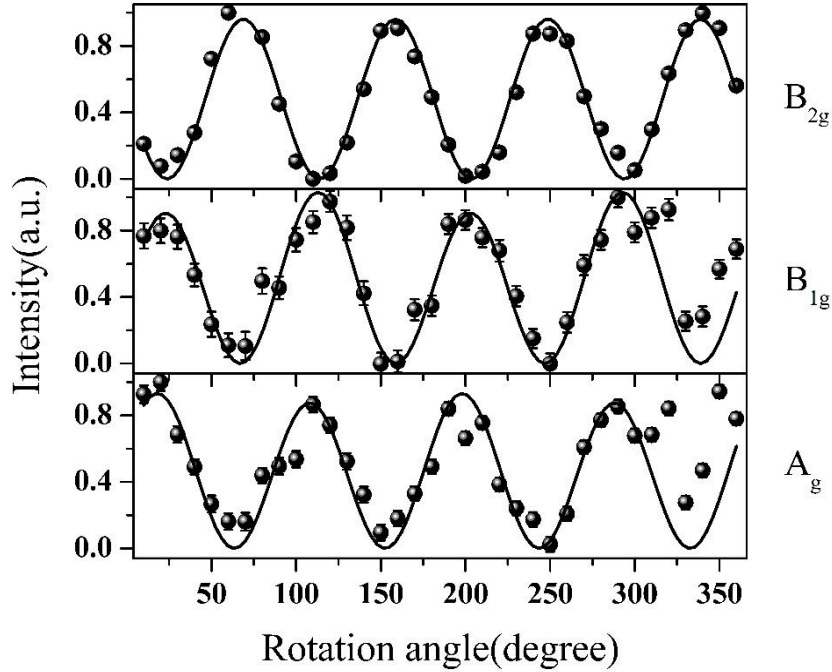

**Fig 3:** Intensity variation of the Raman modes measured in YY configuration as a function of in-plane rotation angle ($\psi$). Solid lines represent theoretical fit using equation 3 and 4.

The room temperature Raman spectra collected in parallel (YY) and cross (YX) polarization configuration on CuO epitaxial film with fixed $\psi$ is shown in figure 2. Appearance of three peaks ($A_g + 2B_g$), their position and polarization dependence matched with previous reports (*12*) confirming monoclinic phase of CuO (*10*). The profile of the Raman spectra are fitted using a Lorentzian function and the intensity and width of the peaks were obtained. The film was rotated along (111) direction i.e. by varying in-plane rotation angle ($\psi$) and the Raman spectra were collected in both YY and YX configuration. The obtained intensity as a function of rotation angle ($\psi$) for all the three modes in parallel polarization (and cross polarization not shown here for brevity) was first normalized with maximum value obtained in 0-360⁰ and then plotted in figure 3. The obtained curves are fitted using equation 3-6. The curves fitted well with the theoretical calculations establishing perfect epitaxial nature of the film. From this, the directions of the crystallographic axes of the CuO epitaxial film is deduced and the obtained results showed concurrence with the φ-scan. From the fitting parameters and using equation 3-6, the values of the Raman tensor elements were deduced (table S2 of the supplementary material). It is seen that the value of the off diagonal elements of the $A_g$ mode Raman tensor is very small when compared to the diagonal terms, this indicate that the $A_g$ mode is highly polarized along the principle axis of the polarizability ellipsoid. This is in consonance with the previous reports, where theoretical calculations suggested that the $A_g$ mode represent vibrations along *b*-axis (*13*). Similarly, the $B_{1g}$ mode arises due to vibrations of the oxygen atoms along *a*-axis while $B_{2g}$ mode represents vibrations perpendicular to both *a*- and *b*-axis (*13*). Thus, $A_g$ mode is ideally suited for observing the displacement along the *b*-axis or in other words, if displacement is along the *b*-axis it will be strongly reflected in the intensity and position of the $A_g$ mode.

Therefore, when the polarization of the incident beam is in the direction of the *b*-axis ($\psi=0°$) and the scattered beam is measured in parallel polarization geometry then in such a configuration only first term in equation (3) will contribute towards the intensity of the $A_g$ mode.

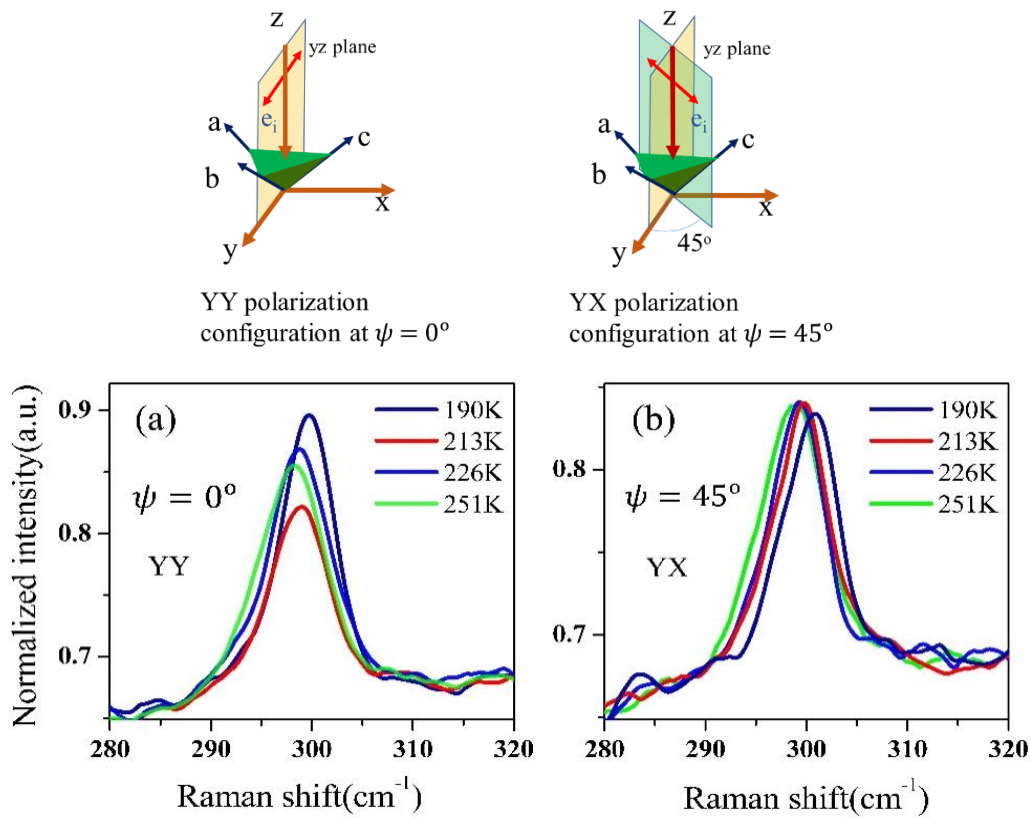

**Fig 4:** Raman spectra at some selected temperatures. (a) $\psi = 0°$ in YY configuration and (b) $\psi = 45°$ in YX configuration. In the top panel, the relation of the crystallographic axes with the laboratory frame of axes is given for the two configurations.

Thus, we argue that when the temperature dependent measurements are performed in this configuration ($\psi=0°$), $I_{A_g}^{\parallel}$ is expected to modulate only if *a* and/or *b* Raman tensor element(s) show temperature dependence. Similarly at $\psi=0°$, the variation of $I_{B_g}^{\parallel}$ with temperature will be decided by the Raman tensor element *e*. Raman tensor element *e* is an off diagonal term and hence the effect of modulation along *b*-axis to the $I_{B_g}^{\parallel}$ (see equation 4) should be relatively small. In the same way, when $\psi=45°$, the *a* and *c* Raman tensor elements are expected to dominate the $I_{A_g}^\perp$ (see equation 5) and the effect of modulation in the value of Raman tensor element *b* is negligible (~ 3% of the total intensity). Equally, when $\psi=45°$, the $I_{B_g}^\perp$ (equation 6) will be decided by the Raman tensor elements *e* and *f*, and both being off-diagonal elements should be less effective in the modulations along the axis of the unit cell.

In order to realize the above mentioned configurations and obtaining temperature dependence of $I_{A_g}^\parallel$ (equation 3) and $I_{A_g}^\perp$ (equation 5), the temperature dependent Raman spectroscopy studies are carried out and presented in figure 4 (a) and 4 (b) in YY ($\psi = 0°$) and YX ($\psi = 45°$) configuration, respectively. The $I_{A_g}^\parallel$ (figure 4(a)) showed a dip at 213 K while $I_{A_g}^\perp$ (figure 4(b)) showed a monotonous behavior with temperature.

The normalized intensities of $A_g$, $B_{1g}$ and $B_{2g}$ modes collected in above mentioned configuration as a function of temperature is plotted in figure 5. Thermal population contributions in the Raman intensities were removed using Bose-Einstein correction factor (*14*). The $I_{A_g}^\parallel$ shows anomalous variation in the incommensurate spin ordered state (AFM2-shaded cyan region). Similarly, the $I_{B_{1g}}^\parallel$ and $I_{B_{2g}}^\parallel$ also showed variation AFM2 region, however, the variations are small when compared to that observed in the $I_{A_g}^\parallel$ (as expected). Further, the $I_{A_g}^\perp$, $I_{B_{1g}}^\perp$, and $I_{B_{2g}}^\perp$ did not show any variation in the entire temperature range. If one observes equation 3 for $\psi=0°$, the $I_{A_g}^\parallel$ is directly proportional to the Raman tensor element *a* and *b*. Thus, large variation in $I_{A_g}^\parallel$ in the incommensurate ordered state is related with variation in Raman tensor element *a* and/or *b*. However, according to equation 5 for $\psi=45°$, $I_{A_g}^\perp$ is directly proportional to Raman tensor elements *a* and *c*, while the contribution due to Raman tensor element *b* is insignificant because of extremely small value of its coefficient (0.04). The variation of $I_{A_g}^\perp$ within error bars in the entire temperature range thus indicates that the Raman tensor elements *a* and *c* are not modulating with temperature.

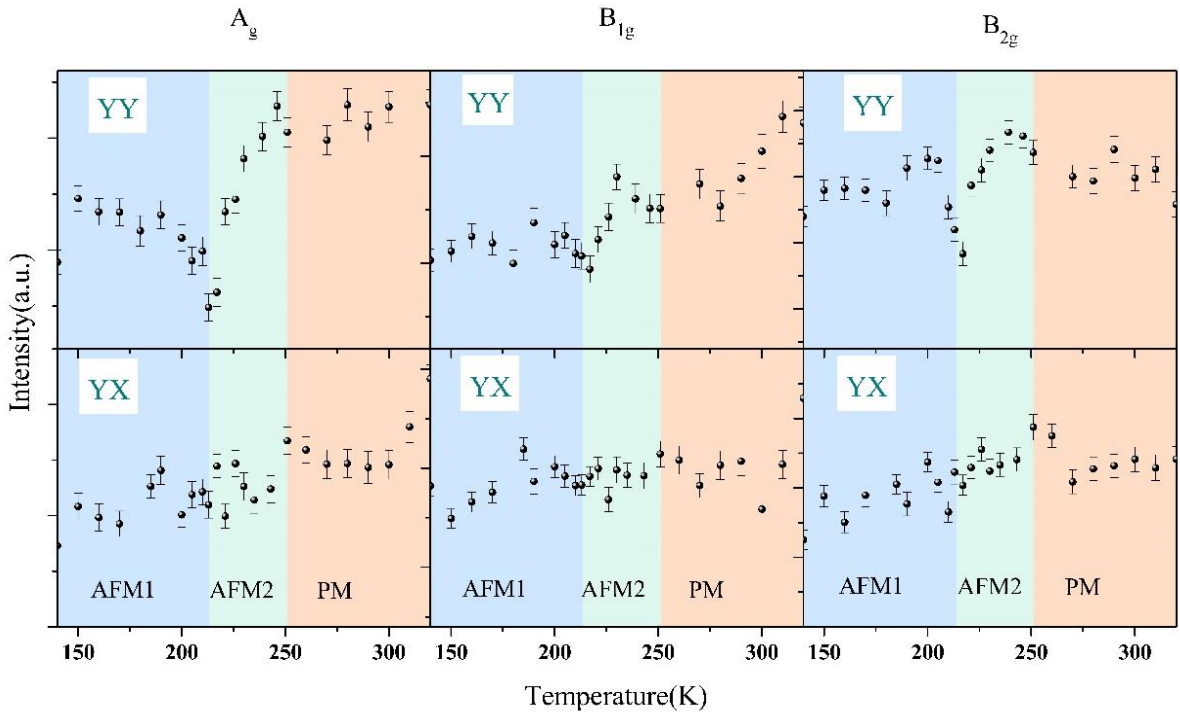

**Fig 5:** Temperature dependence of the intensity of the Raman modes collected using ARPRS at $\psi = 0^0$ (upper) in parallel polarization and at $\psi = 45^0$ (bottom) in cross polarization configuration.

Thus, the large variation shown by $I^{\parallel}_{A_g}$ in the incommensurate magnetic order can be attributed to the variation in Raman tensor element *b*. This proves that the Raman tensor element *b* is only getting modulated due to incommensurate magnetic order. The variation in $I^{\parallel}_{B_g}$ by small amount is due to variation in off-diagonal Raman tensor elements which have components along the *b*-axis. Thus, the temperature dependent polarized Raman spectroscopy study provides a direct evidence of modulation of polarizability ellipsoid along *b*-axis. The polarizability along *b*-axis is directly related to the Cu-O bond distance along b-axis. Thus, this study establishes ionic displacement along *b*-axis due to incommensurate spin order. In order to check the validity of this method, a detailed temperature dependent ARPRS studies are carried out on commercially obtained single crystal substrate of $BaTiO_3$, where the directions of the ionic displacements across the phase transitions is well known. The same has been given in the supplementary material (*7*).

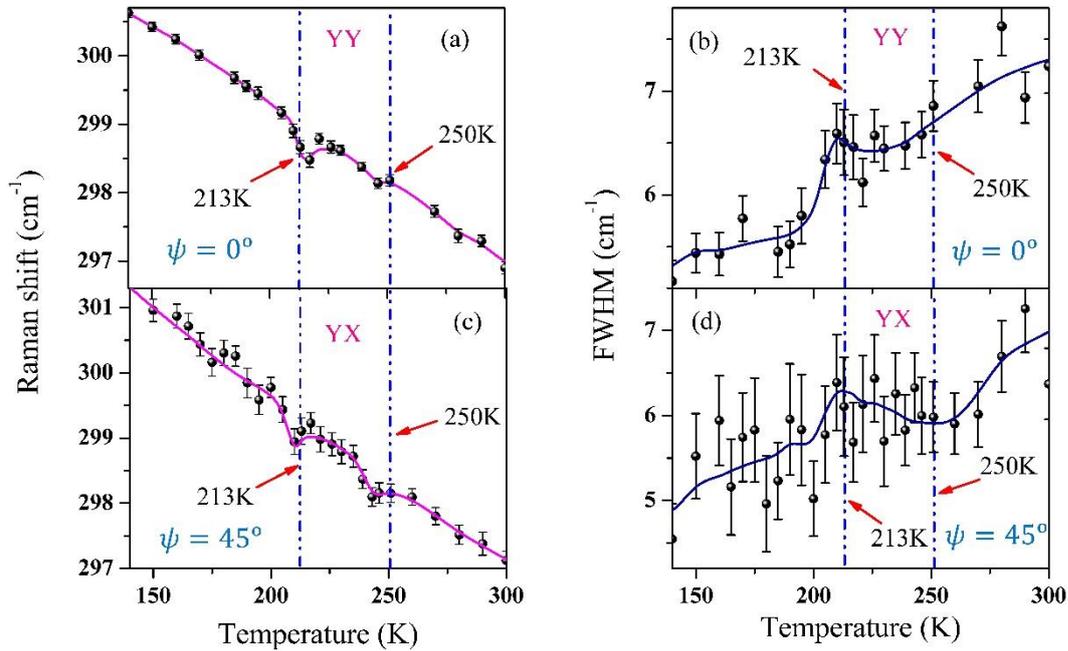

**Fig 6:** Temperature evolution of $A_g$ mode frequency at $\psi=0°$ in parallel-polarization configuration and at $\psi=45°$ in cross-polarization configuration (a & c) and corresponding FWHM (b & d).

In order to quantify the displacement along the *b*-axis due to incommensurate spin order, the Raman shift of the $A_g$ mode as a function of temperature in YY and YX configuration is plotted in figure 6 (a) and (c) respectively. The corresponding FWHM is plotted in figure 6 (b) and (d). Anomalous ionic displacements in the incommensurate state is expected to strongly affect the phonon life time which is reflected in the form of a sharp variation in FWHM of $A_g$ mode around $T_{N1}$. The Raman shift showed a distinct change around 213 K ($T_{N1}$) along with noticeable change around 250 K ($T_{N2}$). The FWHM also showed dramatic variation around this temperature again confirming the involvement of lattice or presence of strong spin-lattice coupling (*13*). The slope of the Raman shift as a function of temperature is obtained by fitting a linear curve and the difference from the linear curve at 213 K (Δω) is obtained. From Δω and using mode Gruneisen parameter (*15*) for the Ag mode, the displacement of the oxygen is calculated which resulted in a value of 13±3 ×10⁻⁵ Å (The details of the calculations are given in supplementary material (*7*)).

The variation of the Raman intensity with temperature is proportional to the variation in the derivative of polarizability (*16*) i.e. variation in Raman tensor elements with temperature. Figure 5 shows that the intensity of the $A_g$ mode ($\psi=0°$) is nearly constant with decreasing temperature before showing a sharp linear decrease with decreasing temperature in the incommensurate spin state region while it shows a gradual increase in the commensurate spin state region. The theoretical calculations by Jin *et al* (*4*) predicted that the spiral spin order in the incommensurate spin state displaces the O atoms along the crystallographic *b*-axis and Cu atoms along the –*b*-axis through inverse D-M interaction and the displacement is very small (~$10^{-5}$Å). Anisotropic exchange mediated displacement of O and Cu atoms in opposite direction produces ferroelectricity along crystallographic *b*-axis which reduces the orbital overlap along *b*-axis. The reduction of covalence of Cu-O bond along *b*-axis reduces the magnitude of the Raman tensor element *b* (polarizability derivative along *b*-axis) resulting in a decrease in the Raman intensity below $T_{N2}$. On further lowering of the temperature, ferroelectricity was reported to increase (*2*) which points towards further increase in the ionic displacement which is reflected in the gradual decrease in Raman mode intensity with decrease in temperature until $T_{N1}$. Below $T_{N1}$ the spin order becomes commensurate resulting in disappearance of D-M interaction along with the associated lattice displacements, non-centrosymmetry and ferroelectricity. In the commensurate spin order state isotropic spin-lattice interaction comes into play (*4*) reducing the Cu-O bond length which increases the covalence resulting in an increase in the Raman intensity below $T_{N1}$. The displacement of Oxygen ion along *b*-axis from its mean position starts below $T_{N2}$ and approaches a maximum value before collapsing at $T_{N1}$. This displacement reduces the Cu-O bond strength (force constant between Cu-O) along *b*-axis that directly affects the Raman shift of the $A_g$ mode. Reimann *et al* (*15*) provided direct evidence of change in frequency of the Ag Raman mode due to oxygen displacement along the *b*-axis through their pressure dependent x-ray diffraction and Raman spectroscopy study. The obtained displacement of the oxygen due to spiral spin state in our study at $T_{N1}$ = 13$\pm$3 $\times 10^{-5}$ Å which is comparable to the theoretically calculated value of 7 $\times 10^{-5}$ Å (*4*). Assuming Oxygen to be in $2^-$ state, the electrical polarization value corresponding to the obtained displacement is ~ 218 μC/m$^2$, which is comparable to the experimentally observed value of 160 μC/m$^2$ (*2*). The small discrepancies in the two values may be due to the deviation in the effective charge of Cu and O ions from $2^{+/-}$.

**Conclusions:**

In conclusion, epitaxially grown thin film of CuO on MgO substrate showed enhancement in the incommensurate magnetic ordering temperature from 230 K to 252 K due to in-plane compressive strain. The value of the Raman Tensor elements were deduced from the detailed polarization dependent Raman spectroscopy study on the grown CuO film. Using the values of Raman tensor elements and temperature dependence of the polarized Raman intensities, the oxygen displacement is confirmed experimentally to be along the *b*-axis. The displacement of the oxygen ions along b-axis is quantified using temperature evolution of the Raman shift in polarized geometry which resulted in a value of 13$\pm$3x10$^{-15}$ meters. The electrical polarization value corresponding to the obtained displacement is ~ 218 μC/m$^2$ against the experimentally reported value of 160 μC/m$^2$. The displacement direction and amplitude agrees with the theoretical prediction and thus provide a direct evidence of the origin of the ferroelectricity due to incommensurate magnetic order in CuO to be because of ionic displacements. This method can be used for obtaining directional displacement at femto length scale in any crystal lattice and in other magnetically induced multiferroic materials in which ionic displacement is the cause of multiferroicity.

**Acknowledgments:**
The authors acknowledge Dr. V.R. Reddy for x-ray diffraction measurements and Dr. R.J. Choudhary for magnetization measurements.

## Materials and Methods

CuO thin film was deposited on MgO (001) substrate by pulse laser deposition. Commercially procured CuO (99.99% from sigma Aldrich) was pelletized and sintered at 950 ⁰C for 8 hours. This sintered pellet was used as a target for deposition. The Substrate was first cleaned for five minutes in acetone and five minutes in methanol using ultrasonic bath. The Film was deposited by using 248nm KrF excimar Laser operated at 220 mJ energy and 4 Hz repetition rate. The chamber was evacuated to high vacuum conditions (~$10^{-6}$ mbar) while during deposition, the substrate was kept at a temperature of 520 ⁰C and oxygen background pressure of 270 mtorr was maintained. The target was rotated at a speed of 10 rpm to prevent pit formation and to ensure uniform ablation of the target. After the deposition, the film was annealed for 5 minutes under the same oxygen partial pressure and then cooled to room temperature with a cooling rate of 20 ⁰C/minute in high oxygen pressure. X-ray diffraction characterization and φ-scan of this film were carried out using PANalytical X'PERT high resolution x-ray diffraction (HRXRD) system equipped with a Cu anode. Polarized Raman spectra was measured using a Horiba Jobin yvon, France make micro-Raman spectrometer equipped with a 473 nm excitation laser, 1800 g/mm grating, edge filter for Rayleigh line rejection and a Peltier cooled CCD detector. The Laser was focused onto the sample with a spot size of ∼ 1 micrometre by a 50× lens. The overall spectral resolution of the system is ∼ 1 $cm^{-1}$ while the uncertainty of the measured Raman peak position was controlled within 0.01 $cm^{-1}$. To obtain polarized Raman spectra, sample was kept on a homemade rotatable sample stage with 0-360⁰ angle marker. Polarized Raman spectra was recorded in parallel (YY) and cross polarization (YX) configuration. In Temperature dependent ARPRS measurements, sample was kept into a Linkam, U.K. make THMS600 stage with a temperature stability of $\pm$ 0.1K. Magnetisation measurement was done using Quantum Design make SQUID –VSM set-up.

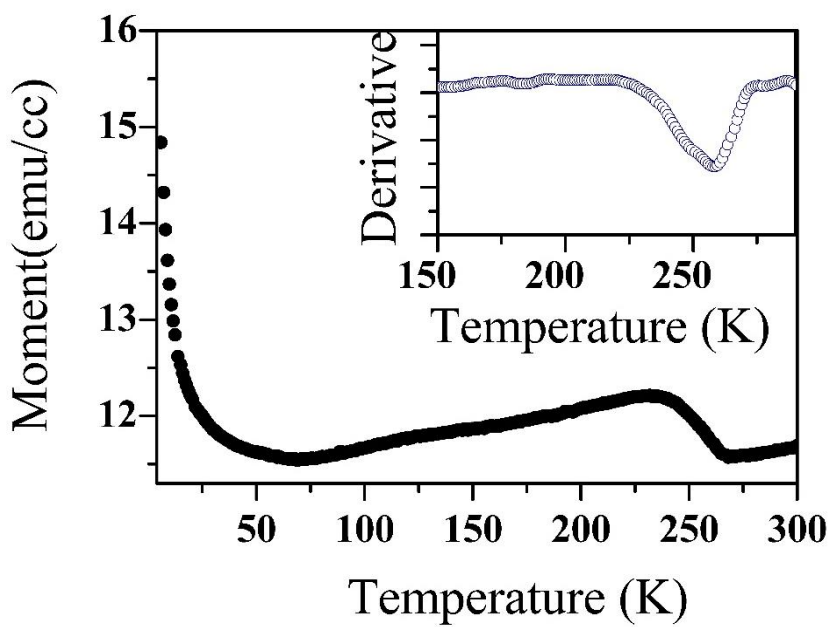

**Fig. S1:** Magnetization as a function of temperature (M-T) measured in Zero Field Cool protocol in the presence of 200 Oe external magnetic field. Inset shows dM/dT.

## Supplementary Text
**Film characterization:**
**1.0 X-ray diffraction studies:**

**1.1 θ-2θ x-ray diffraction**:

The θ-2θ x-ray diffraction scan of the film (Figure 1(a)) showed only two peaks, one at 38.61º, corresponding to monoclinic CuO (111) reflection and the other one at 42.93º corresponding to the (100) reflection of MgO substrate. The CuO (111) peak showed a shift towards lower 2θ value when compared to the bulk (38.79º). Corresponding inter planer spacing along CuO (111) is found to be 2.330 Å that is larger than the bulk value of 2.319 Å indicating out-of-plane tensile strain as reported previously (*17*).

**1.2 Phi scan:**

To obtained the in-plane orientation of this film we have carried out 360º phi scan using CuO($\bar{1}$11) reflection (figure 1(b) ) with a χ value of 62.93º. Sharp peaks indicate that the film has definite in-plane orientation.

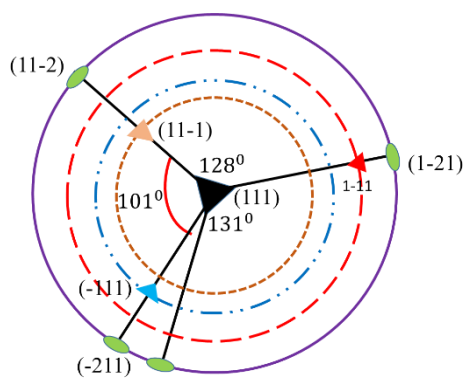

**Fig. S2:** Schematic projection of different planes on (111) plane.

Figure S2 represent stereographic projection with (111) direction as pole. Zone axes of (111) and [$\bar{1}$11] planes is [0$\bar{1}$1] orientation along the plane of the film. The projection of [$\bar{1}$11] on [111] is [$\bar{2}$11] which is normal to both [0$\bar{1}$1] and (111) direction. According to literature MgO (110) matches with CuO ($\bar{1}$10) therefore the normal to these two directions will also be essentially parallel to each other. Now normal to the ($\bar{1}$10) and (0$\bar{1}$1) are (11$\bar{2}$) and ($\bar{2}$11) respectively. In present system the angle between (11$\bar{2}$) and ($\bar{2}$11) directions is 100.84°. In monoclinic system (111) plane have one fold degeneracy. So the phi values of ($\bar{1}$11) diffraction can be given as

$$\Phi = m * 90^0 \pm 100.84^0 + \Phi_0$$

The effect of four fold symmetry of MgO [110] is represented by the first term in the expression while the second term arises from the angle between MgO (110) and CuO ($\bar{1}$10). $\Phi_0$ is the initial phi value (in our case this value was found to be 45° obtained from the phi scan carried out at (111) reflection of the substrate). Experimentally the separation between two peaks was found to be 24.5° as reported in literature (*18*). Observed separation is slightly greater than the expected value from theoretical calculations (21.68°). This discrepancy is attributed to strain effect and is discussed later.

To confirm the growth direction of the film, we also performed φ-scan on (202) reflection (from 0 to 180°). It is seen from the stereographic projection that the projection of (202) on (111) is along (1$\bar{2}$1) that makes ~128° to the (11$\bar{2}$). So the φ values of (202) diffraction spot can be given as

$$\Phi = m * 90^0 \pm 127.93^0 + \Phi_0$$

In this case the φ of (202) substrate reflection, $\Phi_0 = 0°$. Following the arguments presented for the analysis of the φ-scan of ($\bar{1}$11), the angular separation between two successive peaks should be ~14.14° which is also smaller than the experimentally observed separation of ~16.5°. This deviation between experimental and calculated values can be explained by considering strain present in the thin film. The amount of strain present in the system is obtained from Raman measurements that is presented below.

**Strain calculation from Raman shift:**

The direction of vibration for the modes in Raman analysis is obtained from reference (*13*). The table S1 shows minor tensile strain along *b*-axis while significant compressive strain along the *a* and *c*-axes. For bulk sample, the lattice parameters of *a*=4.6837Å, *b*=3.4226Å, *c*= 5.1288Å and *β* =99.54° results in the *d* spacing of (111) plane = 2.322Å. If we consider the modified lattice parameters i.e. after adding Δ*l* from table S1 then the *d* spacing of (111) will be = 2.318Å i.e. *d* spacing decrease. On the contrary the x-ray diffraction showed increase in (111) *d* spacing in thin film when compared to bulk. This is possible only if *β* value reduces in thin film compared to bulk value of 99.54°. On back calculation, the *β* value is found to be 99.24°. Calculations show that the reduction in *a*, *c* and *β* will enhance the angular separation between the peaks in φ-scan as observed in our φ-scan measurements. Reduction in *β* and *a*, *c* lattice parameters should reduce Cu-O-Cu bond distance while the Cu-O-Cu bond angle remains almost constant. The Cu-O-Cu bond is along the ($\bar{1}$01) and hence reduction in Cu-O-Cu bond length should significantly enhance the exchange strength along ($\bar{1}$01) direction. This results in enhancement of T$_{N2}$ to 250K in thin film from the bulk value of 230K.

**Table S1:** Strain calculation from Raman shift.

| modes | ω cm⁻¹ | Δω = ω$_{bulk}$-ω$_{film}$ cm⁻¹ | Mode Grunession parameter Υ | L (lattice parameter) Å | Δl Å | Nature of strain |
|---|---|---|---|---|---|---|
| A$_g$ | 298.24 | 0.23 | 0.86 | b = 3.4226 | 0.0010 | Tensile |
| B$_{1g}$ | 346.11 | -1.97 | 0.66 | a = 4.6837 | -0.0134 | Compressive |
| B$_{2g}$ | 632.06 | -2.11 | 0.34 | c = 5.1288 | -0.0176 | Compressive |

**Polarized Raman calculation:**

In crystalline solids, different phonon branches corresponds to the different vibrational symmetries, and are represented by irreducible representation of the Crystallographic point group (*19*). Hence standard group theoretical calculations give the selection rules of the Raman active modes (*19*). In addition, intensity of polarized Raman spectra depends on the angle between polarization vector of incident light and principle axis of vibration (*20*). By virtue of symmetry, the Principle vibrational axes are defined by the crystallographic axes. Therefore, for a given orientation, the variation in spectral intensity can be calculated using corresponding Raman tensor or in other words if the value of the Raman tensor elements is known, then from the variation of the intensity of the polarized Raman spectra as a function of azimuthal angle, one can get the information about crystallographic orientation.

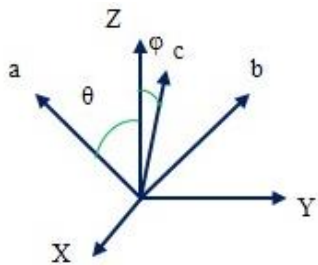

**Fig. S3:** Angular relation between lab axes (XYZ) and crystallographic axes (abc).

According to Porto notation (*21*), polarized Raman spectra is defined by $Z(XY)Z'$, where $Z$ and $Z'$ represent the direction of incident and scattering light respectively and $Y$ and $X$ corresponds to the polarization of electric field of incident and scattered light respectively. According to the group theory, the intensity of the Raman spectra is given as, (*22*)

$$I \propto (e_i R e_s)^2 \tag{1}$$

Where $e_s$ and $e_i$ are the polarization unit vectors of scattered and incident light, respectively while $R$ represents the Raman tensor. In back scattering geometry (present instrumental geometry), for a given incident polarized light, cross and parallel are two possible configurations geometry for collecting scattered light. Following matrix representation the unit vectors of incident and scattered light in various polarization configuration can be represented as:

$$e_i^{\parallel} = (0 \quad 1 \quad 0)$$
$$e_s^{\parallel} = \begin{pmatrix} 0 \\ 1 \\ 0 \end{pmatrix} \tag{2}$$

$$e_s^\perp = \begin{pmatrix} 1 \\ 0 \\ 0 \end{pmatrix}$$

where $i, s$ correspond to incident and scattered light, $\perp$ and $\parallel$ represent cross and parallel polarization configuration of electric field vector, respectively. In order to facilitate calculations, the crystallographic unit cells are essentially required to be transformed to the laboratory frame of references. For the same we used the standard Euler matrix (20) $\Phi$ to transform crystal axes to lab axes and the transformed Raman tensor ($R'$) in Lab axes can be written as follows (20)

$$R' = \Phi R \bar{\Phi} \tag{3}$$

Figure S3 gives the transformation of the crystal axes to lab axes. θ & φ are Euler angles. In present case where the film is grown along (111) direction and the CuO ($1\bar{1}0$) in the plane of the film is along the MgO (110) orientation, the value of θ and φ are ∼ 47.57º and ∼ 53.85º, respectively considering orthogonal crystallographic axes.

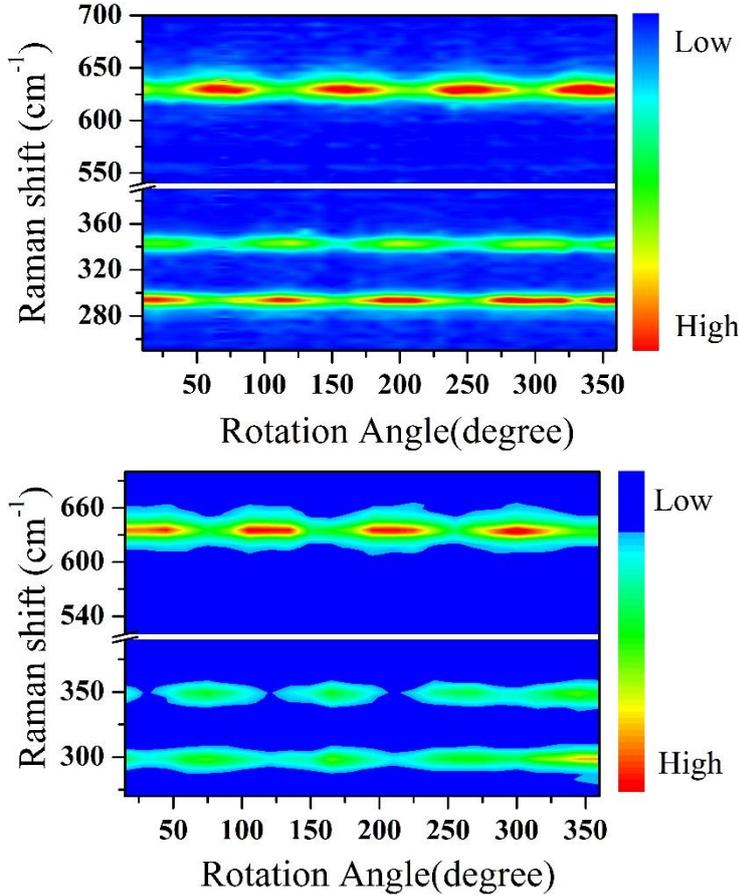

**Fig. S4:** Contour plot of Raman spectra with in plane rotation angles in different polarization configuration YY (upper) and YX (bottom).

From equation 1, 2 and 3, the angular dependence (in plane rotation angle $\psi$) of the intensity of the Raman modes can be calculated as

$$I_{A_g}^{\parallel} \propto \left((0.64a + 0.34b)\cos^2\psi + (0.15a + 0.28b + 0.54c - 0.56d)\sin^2\psi + (0.31(a-b) - 0.58d)\sin2\psi\right)^2 \tag{4}$$

$$I_{B_g}^{\parallel} \propto \left((-0.95e)\cos^2\psi + 2(0.21e - 0.20f)\sin^2\psi + (0.20e + 0.43f)\sin2\psi\right)^2 \tag{5}$$

$$I_{A_g}^{\perp} \propto \left((0.31(b-a) + 0.58d)cos2\psi + \frac{(0.49a+0.04b-0.54c+0.56d)}{2} sin2\psi\right)^2 \quad (6)$$

$$I_{B_g}^{\perp} \propto \left((-0.21e - 0.43f)cos2\psi + (-0.69e + 0.42f)sin2\psi\right)^2 \quad (7)$$

The intensity of the Raman mode is measured as a function of in-plane rotation angle $\psi$ and the same is presented in figure S4 which shows sinusoidal modulations.

**Determination of the Raman tensor elements**:
From fitting the polarization dependent Raman intensity (figure 3), tensor elements follow these equations

$$0.64a + 0.34b = 0.884$$
$$0.15a + 0.28b + 0.54c - 0.56d = -0.854$$
$$0.31(a - b) - 0.58d = -0.385$$
$$0.49a + 0.04b - 0.45c + 0.56d = -1.686$$
$$-0.95e = 0.936$$
$$0.43e - 0.80f = -0.999$$

The normalized values obtained by solving the above equations are tabulated in table S2.

**Table S2:** The normalized Raman tensor elements values.

| Modes | Tensor elements | mean square error |
|---|---|---|
| Ag | a = 0.17 | ± 0.03 |
|    | b = 0.56 | ± 0.06 |
|    | c = -0.84 | ± 0.01 |
|    | d = 0.02 | ± 0.03 |
| Bg | e = -0.33 | ± 0.01 |
|    | f = 0.24 | ± 0.03 |

**Displacement calculation:**
It is well known that the Mode Grunession parameter ($\Upsilon$) links the change in volume to the change in frequency of the mode

$$\Upsilon = -\frac{\Delta ln\omega}{\Delta lnV} = -\frac{\Delta\omega}{\omega} \cdot \frac{l}{3\Delta l}$$

Values of $\Delta l = l\frac{\Delta\omega}{3\Upsilon\omega}$;

$\Delta\omega = 0.36\ cm^{-1}, l = 1.96\ Å\ (bond\ length\ of\ Cu - O\ along\ \bar{1}01\ chain), \Upsilon = 0.86\ for\ Ag\ mode\ (15), \omega = 298.84\ cm^{-1}$.

$\Delta l = 9.15 \times 10^{-4}\ Å$

Displacement of oxygen atom along crystallographic b axis
$= \frac{\Delta l}{2} cos\theta; (\theta \sim 73°$, Cu-O-Cu angle along $\bar{1}01$ chain)
[Oxygen and Cu atoms displaced in opposite direction. Half factor for only oxygen atom displacement]
$\cong (13 \pm 3) \times 10^{-5}\ Å$

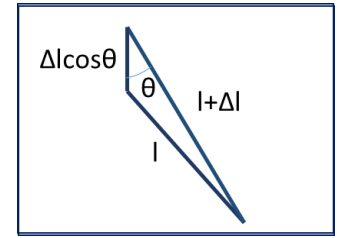

Electric polarization due to O ion displacements
$p = 4Nq\Delta l$;
N – number of unit cell per unit volume (12.33×10²⁷ m⁻³), q - charge of ions (= 2), $\Delta l$ - displacement (13×10⁻⁵ Å), four factor represents unit cell contained four molecules.
$= 218\ \mu c/m^2$

# ARPRS Study of Raman mode intensity across phase transitions in BaTiO3

In this section, we describe the dependence of Raman mode intensity in single crystal of BaTiO3 as a function of temperature and polarization direction. We followed the normal modes and its graphical representation as given by Freire et al (*23*). It is well known that the A1(TO2) mode represents vibration along c-axis and hence is ideal to get information of the displacement of the Ti-ions. The temperature dependence of the intensity of the A1 mode, from 100 K to 370 K, is measured such that the polarization of the incident light is parallel to one of the crystallographic axis. BaTiO3 shows cubic to tetragonal transition at 393 K, tetragonal to orthorhombic transition at 278 K while an orthorhombic to rhombohedral transition at 190 K. As we are presenting the analysis of A1 mode, we give below the Raman tensor of the A1 mode in all the crystallographic structure displayed by BaTiO3 at various temperatures. The measurements were performed on single crystal substrate of BaTiO3 (100) which was procured commercially. It is well known in BaTiO3 single crystal that the *c*- and *a*-axis domains co-exist (*24*). We take the case for the (100) domain (*a*-axis domain). The Raman tensor element for the A1 mode in various symmetries is given as (25, 19):

$$A_1 = \begin{pmatrix} a & 0 & 0 \\ 0 & a & 0 \\ 0 & 0 & b \end{pmatrix} \quad \text{for tetragonal}$$

$$A_1 = \begin{pmatrix} p & 0 & 0 \\ 0 & q & 0 \\ 0 & 0 & r \end{pmatrix} \quad \text{for orthorhombic}$$

$$A_1 = \begin{pmatrix} l & 0 & 0 \\ 0 & l & 0 \\ 0 & 0 & m \end{pmatrix} \quad \text{for rhombohedral}$$

The intensity of the A1 mode is calculated by using the method described for CuO in previous section. The resulted relations for the intensity of A1 mode in (100) oriented domains in parallel polarization configuration are given as:

$$I_{A_1}^{\parallel} = [a\cos^2\psi + b\sin^2\psi]^2 \quad \text{------------------For tetragonal phase} \quad (8)$$

$$I_{A_1}^{\parallel} = \left[\left(\frac{p}{2}\right)\sin^2\psi + q\cos^2\psi + \left(\frac{r}{2}\right)\sin^2\psi\right]^2 \quad \text{---For orthorhombic phase} \quad (9)$$

$$I_{A_1}^{\parallel} = [l\cos^2\psi + m\sin^2\psi]^2 \quad \text{------------------For rhombohedral phase} \quad (10)$$

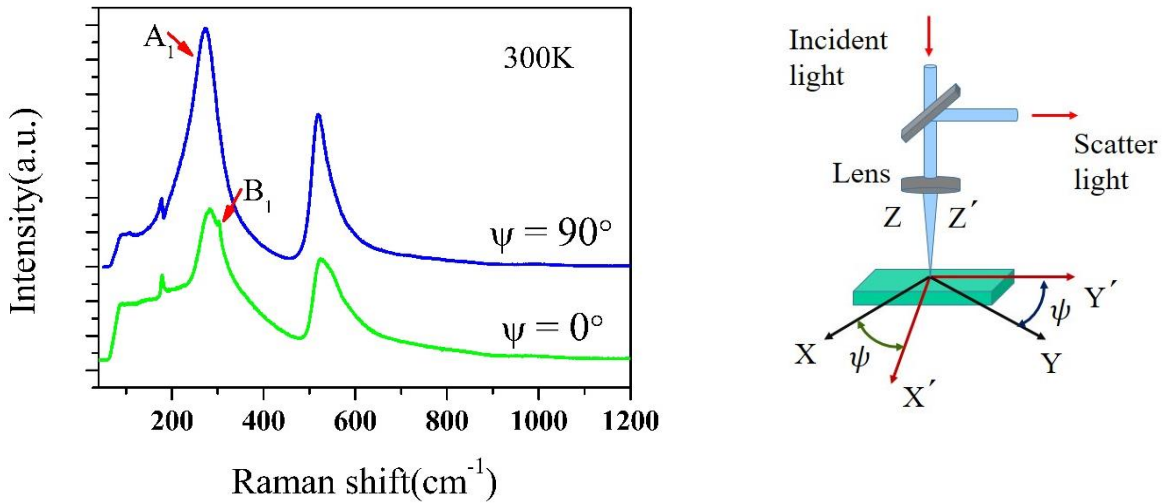

**Fig. S5:** Raman spectra collected at 300 K in two different in-plane direction ($\psi=0°$ and $\psi=90°$) in parallel polarization configuration. The schematic on the right hand shows the experimental geometry.

It is well known that in tetragonal phase the Ti atoms show displacement along the c-axis. The A1 mode represents displacement of all the atoms along the c-axis and therefore it can be effectively used to track the variation in intensity due to displacement of Ti atoms in the unit cell of $BaTiO_3$. We measured the Raman spectra after identifying (100) and (001) oriented domain regions in parallel polarization configuration that are shown in figure S5. The temperature dependence of A1 mode intensity collected in parallel polarization configuration thus will be guided by the equations given above in various temperature regions. We had collected the data in two configurations.

**Configuration (1): $\psi = 90°$ (c-axis parallel to polarization)**

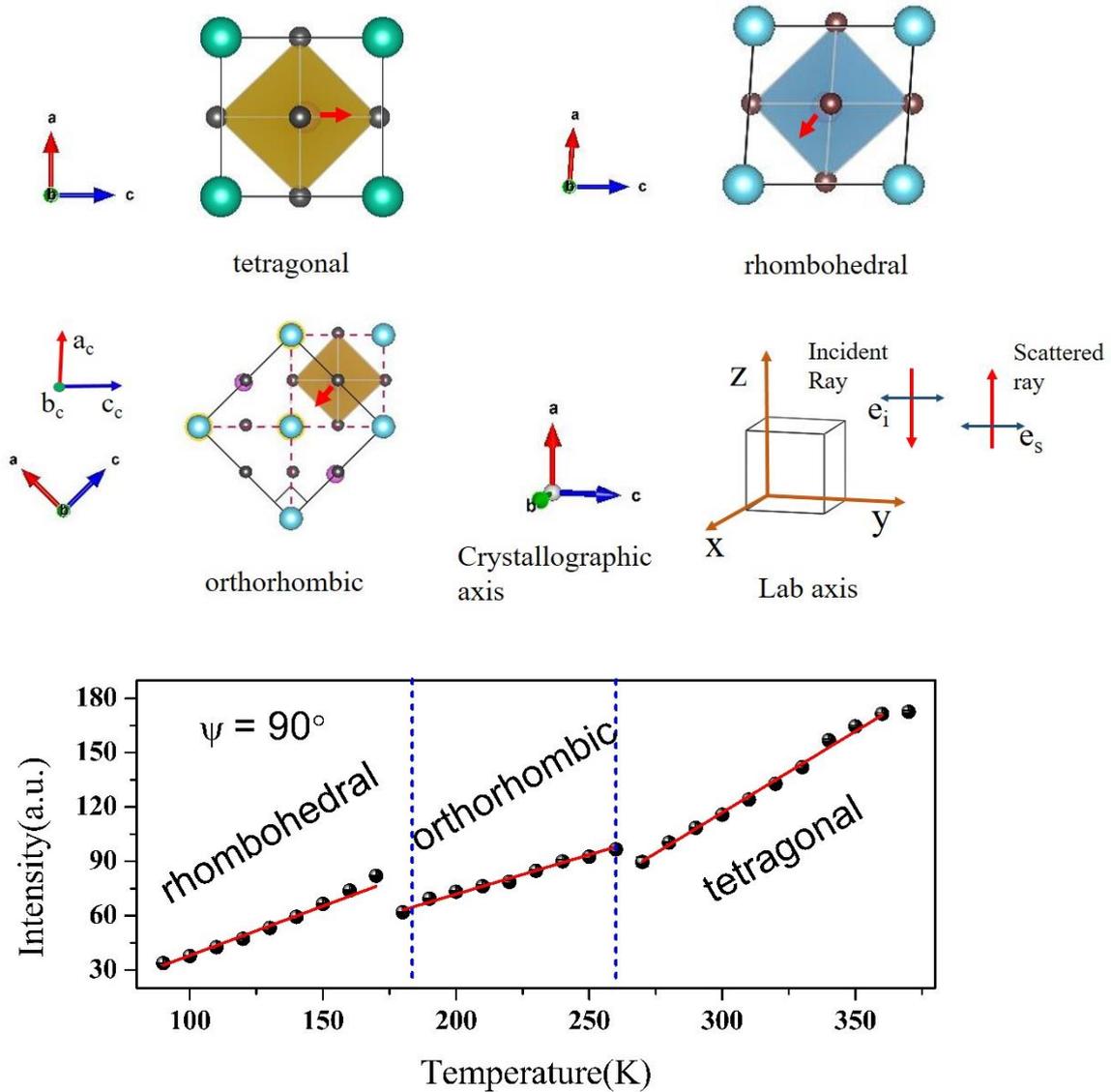

**Fig. S6:** The intensity of the Raman mode A1 as a function of temperature collected in parallel polarization configuration with $\psi = 90°$. The dots represent the experimental points while the line is a guide to the eye. The top panel shows the unit cell of the $BaTiO_3$ in tetragonal, orthorhombic and rhombohedral symmetry along with unit cell directions and measurement geometry. The arrow (red) shows the displacement direction of the Ti.

If we put $\psi = 90°$ in equations (8-10), only element *b* of the Raman tensor in the tetragonal (equation 8) and element *p* and *r* in the orthorhombic phase (equation 9) will contribute to the intensity of the A1 mode. In tetragonal phase Ti-ions show displacement along crystallographic

*c*-axis which increases with decreasing temperature. This should change the value of the Raman tensor element '*b*' and therefore when data is collected with $\psi = 90^o$ in parallel polarization configuration, the intensity is expected to show a variation with temperature. Figure S6 shows the intensity as function of temperature of A1 mode collected in this configuration. It showed a linear decrease in intensity with decrease in temperature. Similarly, in case of orthorhombic phase, Ti-ions displaces along (001) orthorhombic unit cell direction. This ionic displacement thus expected to change only the value of the Raman tensor element *r*. Thus, the intensity of the A1 mode is expected to change with change in temperature as shown experimentally in figure S6. The intensity of the A1 mode was determined by fitting the peak at ~270 cm$^{-1}$ by using a Lorentzian function after correcting the spectra for Bose-Einstein thermal factor. It is noted that the slope of the decrease in intensity with decrease in temperature in orthorhombic phase is significantly less when compared to that in the tetragonal phase. This again proves the involvement of the Raman tensor elements in identifying the displacement direction of the cation. In case of orthorhombic phase, the displacement is along the (001) direction of the orthorhombic unit cell which signifies the (101) direction in the laboratory frame of reference (see figure S6 top panel). Therefore, the projection of the Raman tensor element '*r*' (which is 45° from z-axis of the laboratory frame of axes) on the z-axis of the laboratory frame of axes will only contribute in the intensity of this mode and hence the temperature dependence of the intensity is expected to be 2 times lesser than that in the tetragonal phase, if we consider the values of the Raman tensor element *b* and *r* to be same. We calculated the slope in orthorhombic phase and found it to be 0.44 while in tetragonal phase it is 0.89. Thus, it matches with the theoretical expectations. In rhombohedral phase, Ti-ions displaces along the (111) axis, so all the tensor elements are expected to show change with temperature which is reflected in decrease in intensity with decreasing temperature (figure S6). It is worth noting here that we have observed a clear discontinuity in the intensity vs temperature plot of the A1 mode measured in cooling cycle across the phase transition temperatures. Also it is worth noting here that the polarization in BaTiO$_3$ is reported to show a nearly linear increase with decreasing temperature (*26*). Thus, the A1 mode intensity in this configuration is a direct reflection of the spontaneous polarization component in BaTiO$_3$.

**Configuration (2): $\psi = 0^o$ (b-axis parallel to polarization)**

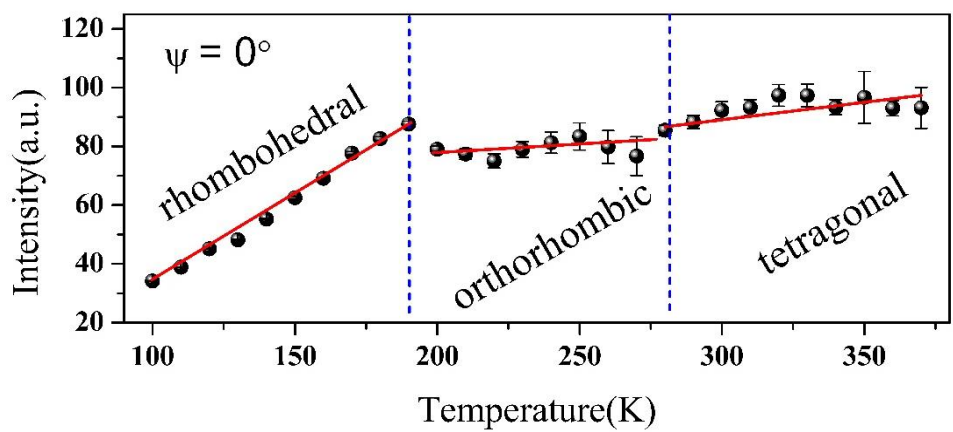

**Fig. S7:** The intensity of the Raman mode A1 as a function of temperature collected in parallel polarization configuration with $\psi = 0^o$. The dots represent the experimental points while the line is a guide to the eye.

If we put the value of $\psi = 0°$ in equation (8) i.e. for the tetragonal phase, it suggests that only Raman tensor element *a* will contribute to the intensity. Similarly equation (9) i.e. for the orthorhombic phase, suggests that only Raman tensor element *q* will contribute to the intensity of A1 mode. As mentioned before in tetragonal and in orthorhombic phase Ti-ions displace along *c*-axis in the corresponding unit cell. Therefore, the Raman tensor elements *a* and *q* should be invariant with temperature. Figure S7 shows the intensity of the Raman mode collected in this configuration. It showed very weak temperature dependence in the tetragonal phase while it is temperature independent in the orthorhombic phase. In tetragonal phase also the intensity of the Raman mode should have been constant, the very small decrease in intensity with decrease in temperature is attributed to the tilt of the domains with respect to the out of plane direction as reported previously by many workers (*24*). In rhombohedral phase i.e. equation (10) the Raman tensor element *l* is expected to contribute to the intensity of the A1 mode. As the displacement of the cation is along the (111) direction, the Raman tensor element *l* is expected to change with change in temperature and thus the intensity of the A1 mode (figure S7).

In conclusion, it is shown in single crystal $BaTiO_3$ that the intensity of a Raman mode is proportional to the spontaneous polarization along a particular direction which is directly related with the cationic displacement in that direction. Hence the Raman intensity if collected in directional manner (ARPRS) can track the smallest cationic displacement. We have used this aspect in tracking the exceedingly small displacement of the oxygen ion in CuO which is presented in main text.